\documentclass{aa}
\usepackage{epsfig}
\usepackage{graphicx}
\usepackage{mathtools}
\usepackage{txfonts}
\usepackage{lscape}
\usepackage{hyperref}
\usepackage{color}
\usepackage{amsmath}

\bibpunct{(}{)}{;}{a}{}{,}

\begin{document}
\title{
Detection of the Milky Way spiral arms in dust from 3D mapping
}
\subtitle{}

\author{Sara Rezaei Kh.\inst{\ref{inst1}} \and Coryn A.L. Bailer-Jones\inst{\ref{inst1}} \and David W. Hogg\inst{\ref{inst1}\ref{inst2}\ref{inst3}\ref{inst4}} \and Mathias Schultheis\inst{\ref{inst5}}\\
}
\institute{Max Plank Institute for Astronomy (MPIA),  K\"onigstuhl 17, 69117 Heidelberg, Germany \label{inst1} 
\and 
Center for Computational Astrophysics, Flatiron Institute, 162 Fifth Ave, New York, NY 10010, USA \label{inst2} 
\and 
Center for Cosmology and Particle Physics, Department of Physics, New York University, 726 Broadway, New York, NY 10003, USA \label{inst3} 
\and 
Center for Data Science, New York University, 60 Fifth Ave, New York, NY 10011, USA \label{inst4} 
\and
Laboratoire Lagrange, Universit\'e C\^ote d'Azur, Observatoire de la C\^ote d'Azur, CNRS, Blvd de l'Observatoire, F-06304 Nice, France \label{inst5}
}

\def\teff{$T_{\rm eff}$}

\def\expec{{\mathrm E}}
\def\cov{{\mathrm Cov}}

\def\trans{^\mathsf{T}}
\def\inv{^{-1}}
\def\mby{\!\times\!}
\def\mplus{\!+\!}

\def\geonj{g_{n,j}}
\def\gvec{{\mathbf g}}
\def\geomat{{\mathrm G}}

\def\likecovN{{\mathrm V}_N}

\def\extn{a_n}
\def\extsdn{\sigma_n}
\def\extvecN{{\mathbf a}_N}
\def\extcovN{\Sigma_N}
\def\estextn{f_n}
\def\estextvecN{{\mathbf f}\!_N}

\def\rhonj{\rho_{n,j}}
\def\rhovecJ{{\boldsymbol \rho}_J}
\def\rhovecJp{{\boldsymbol x}_{N+1}}
\def\rhoJp{\rho_{new}}

\def\gpcov{{\mathrm C}}
\def\gpcovJ{{\mathrm C}_J}
\def\gpcovJp{{\mathrm \Omega}_{N+1}}
\def\gpcovel{c}
\def\kvecJ{{\mathbf k}_J}
\def\kJp{k}

\def\mmatJ{{\mathrm Q}_N}
\def\mvecJ{{\mathbf q}_N}

\def\bvecJ{{\mathbf b}_N}
\def\hvecJ{{\mathbf h}_N}
\def\rmatJ{{\mathrm R}_N}

\def\rvec{{\mathbf r}}
\def\ofo{{\mathcal O}}

\def\meanrho{\vec {\rho}_{\mu}}


\abstract{
Large stellar surveys are sensitive to interstellar dust through the effects of reddening. Using extinctions measured from photometry and spectroscopy, together with three-dimensional (3D) positions of individual stars, it is possible to construct a three-dimensional dust map. We present the first continuous map of the dust distribution in the Galactic disk out to 7 kpc within 100 pc of the Galactic midplane, using red clump and giant stars from SDSS APOGEE DR14. We use a non-parametric method based on Gaussian Processes to map the dust density, which is the local property of the ISM rather than an integrated quantity. This method models the dust correlation between points in 3D space and can capture arbitrary variations, unconstrained by a pre-specified functional form. This produces a continuous map without line-of-sight artefacts. Our resulting map traces some features of the local Galactic spiral arms, even though the model contains no prior suggestion of spiral arms, nor any underlying model for the Galactic structure. This is the first time that such evident arm structures have been captured by a dust density map in the Milky Way. Our resulting map also traces some of the known giant molecular clouds in the Galaxy and puts some constraints on their distances, some of which were hitherto relatively uncertain.
}


\maketitle

\section{Introduction}

Attempts to map our Milky Way date back to the 18th century. One of the most important works was by William Herschel, who constructed a map of the Milky Way by counting stars in more than 600 different lines of sight. He concluded that the Milky Way is a flattened disk and the Sun is located very close to the centre \citep{Herschel75}. Jacobus Kapteyn improved Herschel's map using photometric star counts and parallaxes and proper motions of stars and estimated the size and shape of the Milky Way \citep{Kapteyn22}. However, neither Kapteyn nor Herschel were aware of the importance of the extinction of light by interstellar dust, which resulted in erroneous estimates of the size and shape of the Galaxy. Robert Trumpler found the first evidence of the interstellar reddening and demonstrated how significant the interstellar dust extinction is \citep{Trumpler30} that makes distant objects look fainter than they would in the absence of dust.

Recognising the effects of the interstellar dust on the observations, many attempts were made by astronomers to map the extinction in the Milky Way. One of the significant studies in this regard is the work by \citet*{Schlegel98} who mapped the dust column density using far-infrared dust emission from the IRAS and COBE satellites. A more sensitive 2D map with higher resolution was made by \cite{Planck14} using a similar method to \citet*{Schlegel98}. However, for many Galactic studies, 2D measurements do not suffice; we often need an estimate of the three-dimensional location of the emitting/extinguishing sources in the Galaxy. Moreover, dust plays an important role in creating and shaping the Galaxy and forming stars and planets. Knowing the local distribution of dust provides valuable information about the Galactic structure and the probable sites of star formation.

This opened a new area of studies in which various groups have been trying to map the Galactic dust extinction in 3D using different data sets and techniques. \cite{Marshall06} presented a 3D extinction model in the Galactic plane using a Galactic model and the near infrared colour excess to estimate distances and extinctions. \cite{Schlafly10} used the blue tip of the distribution of stellar colours to measure the colour of the main sequence turnoff stars and measured the reddening of stars in the SDSS-III footprint. \cite{Sale12} developed a hierarchical Bayesian model to simultaneously infer extinction and stellar parameters from multi-band photometry, and \cite{Sale14} used this method to build a 3D extinction map of the northern Galactic plane using IPHAS photometry. A similar probabilistic method was developed by \citet*{Hanson14} to estimate the effective temperature and extinction based on a method previously introduced by \citet*{CBJ11}. They used a Bayesian framework to account for the degeneracy between extinction and stellar effective temperature to produce a 3D extinction map of the Galactic high latitudes (b > $\sim30^{\circ}$) using SDSS and UKIDSS. \cite{Hanson16} then used photometry from Pan-STARRS1 and Spitzer Glimpse surveys to map the dust extinction in the Galactic plane. \cite{Green14} introduced a method similar to \cite{Sale12} to determine dust reddening from stellar photometry which was then used by \cite{Schlafly14} to map the dust reddening of the entire sky north of declination $-30^{\circ}$. This was also used later by \cite{Green15} to build a 3D map of dust reddening for three-quarters of the sky using Pan-STARRS1 and 2MASS. \cite{Green18} recently introduced an updated version of the map using a more accurate extinction law and additional new data from Pan-STARRS1.

The main drawback of these methods is that they treat each line of sight (l.o.s) independently from one another. This creates artefacts and discontinuities in their results. \cite{Vergely10} used a method with a smoothing kernel to account for gaps in the data, and mapped the dust opacity (mag/pc) in the Sun's vicinity. A similar approach was taken by \cite{Lallement14} who presented a 3D map of the local opacity. They later updated this map in \citet{capitanio17} using distance information from Gaia TGAS and colour excess estimates from diffuse interstellar bands (DIBs) from SDSS/APOGEE spectra, adopting a low-resolution map based on Pan-STARRS1 reddening measurements as a prior. \citet*{SaleM14} introduced a new method to map the Galactic extinction and dust in which the logarithm of the extinction (log A) is modelled as a Gaussian random field. Its covariance function has a Kolmogorov-like power spectrum which is motivated by a physical model of the interstellar medium. Dust is also modelled as a semi-stationary random field which produces a log A distribution that is very close to Gaussian.

Evidence for spiral structure in the Milky Way dates back to 1951 when W.W. Morgan and collaborators determined the distances towards emission regions \citep{Oort52,Morgan53}. This was confirmed shortly after by the discovery of 21 cm radio observations \citep{Hulst54,Morgan55}. Despite several attempts at 3D dust mapping, none of the current dust maps reveal the Galactic spiral arm structure. This is in contrast with the fact that spiral arms are rich in gas and dust where many stars are formed \citep{Kennicutt11, Schinnerer17}. The reason for this failure of extinction maps is mainly due to the lack of precise distance measurements as well as the assumptions behind dust mapping techniques. Most of the aforementioned maps illustrate dust extinction or reddening which is an integrated property, and so cannot trace local properties of the Galaxy. In addition, many of these methods treat each l.o.s separately such that no information is propagated from neighbouring points, resulting in discontinuities between neighbouring lines of sight in the resulting maps.

We address issues of previous 3D dust extinction maps in our approach by using a non-parametric method to capture complex structures present in the observed data. Furthermore, we directly map dust density - which represents local properties of the Galaxy - rather than the integrated extinction. We take into consideration the correlation between dust points in space using an isotropic Gaussian process that provides a continuous map without l.o.s artefacts. In \citet{Rezaei_Kh_17} we described our approach in detail. We have since improved the method in some ways: these we explain in section \ref{method}. In section \ref{data} we explain our data selection from the Sloan Digital Sky Survey IV's Apache Point Observatory Galactic Evolution Experiment \citep[APOGEE-2,][]{Blanton17, Majewski17, Abolfathi17} which, being an infrared survey, enables to probe to greater distances and/or through dustier regions. We show our map of the Galactic disk in section \ref{map} and discuss the limitations and future possible improvements in section \ref{discussion}.

\section{Method improvements}\label{method}

In \citet{Rezaei_Kh_17} we introduced our non-parametric method to infer the dust distribution in 3D space. The method uses 3D positions of stars and their l.o.s extinctions as the input data, divides the l.o.s towards each star into small 1D cells, then combines all cells in 3D space using a Gaussian process to predict the posterior PDF $P(\rhoJp | \extvecN)$ over the dust density $\rho$ at any point.
In \citet{Rezaei_Kh_17} we defined our Gaussian process prior as
\begin{equation}
P(\rhovecJ) \,=\, \frac{1}{(2\pi)^{J/2}|\gpcovJ|^{1/2}} \exp \left[ -\frac{1}{2} \rhovecJ\trans \gpcovJ\inv \rhovecJ \right] \ ,
\label{eqn:gpJ}
\end{equation}
which is a zero-mean $J$-dimensional Gaussian with the covariance $\gpcovJ$, where $J$ is the total number of all cells towards all $N$ stars in the data ($J\geq N$) and the $J$-dimensional vector $\rhovecJ$ is the set of all $J$ dust densities in all lines of sight. As we discussed in \citet{Rezaei_Kh_17}, this involves a J$\times$J matrix inversion, followed by a set of $J$-dimensional matrix manipulations, which makes the calculations computationally expensive.

Here we introduce a new prior as
\begin{multline}
P(\geomat\rhovecJ) \,=\, \frac{1}{(2\pi)^{J/2}|\geomat\gpcovJ\geomat\trans|^{1/2}} \exp \left[X  \right] \ , \\
X \,=\, -\frac{1}{2} (\geomat\rhovecJ - \geomat\meanrho)\trans (\geomat\gpcovJ\geomat\trans)\inv (\geomat\rhovecJ -\geomat\meanrho) \ .
\label{eqn:gpJn}
\end{multline}
This is still a Gaussian but now in $\geomat\rhovecJ$. This has the advantage of a dramatic drop in its dimensionality from $J$ to $N$, where $N$ is the total number of stars in the sample and $\geomat$ is an $N\mby J$ matrix containing geometric factors of the cells. For our application in this paper $N$ is typically 5\,000 and $J$ about 100\,000. The $J\mby J$ matrix $\gpcovJ$ needs to be calculated once, and $\geomat\gpcovJ\geomat\trans$ is an $N\mby N$ matrix which needs to be inverted also once. Note that the matrix $\gpcovJ$ that is calculated first is built based on the distances between J dust cells, then it is multiplied by the matrix $\geomat$. This leads to a dramatic computational gain, both in terms of speed and the maximum $J$ which can be handled, while providing results identical to using the previous J-dimensional Gaussian process (equation \ref{eqn:gpJ}). For instance, adopting the same $J$ in both cases, the improved method runs about 1000 times faster.
It is also important to note that we allow for a non-zero mean in the Gaussian process prior ($\geomat\meanrho$). As described later, this is determined from a global property of the input data.

Our likelihood function is:
\begin{alignat}{2}
P(\extvecN | \geomat\rhovecJ) \,&=\, \frac{1}{(2\pi)^{N/2}|\likecovN|^{1/2}}  \nonumber\\
   \,&\,\,\quad \mby \,\,\exp\left[ -\frac{1}{2} (\extvecN - \geomat\rhovecJ)\trans \likecovN\inv (\extvecN - \geomat\rhovecJ) \right]  \ ,
\label{eqn:likeNt}
\end{alignat}
where $\extvecN$ is the vector of attenuation measurements with covariance $\likecovN$. The attenuation is unitless and the relation between extinction and attenuation is $A_n \simeq 1.0857 \extn$ \citep{Rezaei_Kh_17}.

The posterior PDF of the dust density at a new point is calculated by multiplying the Gaussian process prior (equation \ref{eqn:gpJn}) by the likelihood (equation \ref{eqn:likeNt}) and marginalising over the $\rhovecJ$ to give
\begin{equation}
P(\rhoJp | \extvecN) \,=\, \sqrt{\frac{\alpha}{2\pi}} \exp \left[   -\frac{\alpha}{2}\left(\rhoJp + \frac{\beta}{\alpha} - {\rho}_{\mu}\right )^2 \right]
\label{eqn:rhointF}
\end{equation}
which is a Gaussian with mean $-\beta/\alpha + \,{\rho}_{\mu}$ and variance $1/\alpha$ where
\begin{alignat}{2}
\alpha \,&=\, q - \mvecJ\trans\rmatJ\inv\mvecJ \nonumber \\
\beta \,&=\, \extvecN\trans\likecovN\inv\rmatJ\inv\mvecJ + \mvecJ\trans\rmatJ\inv\mmatJ\trans\geomat\meanrho - \mvecJ\trans\geomat\meanrho \nonumber \\
\rmatJ \,&=\, \mmatJ + \likecovN\inv
\label{eqn:auxtermsA}
\end{alignat}
and $\mmatJ$ ($N\mby N$ matrix), $\mvecJ$ ($N\mby 1$ vector) and $q$ (scalar) are the partitioned elements of matrix $(\geomat\gpcovJ\geomat\trans)\inv$ (see appendix \ref{sec:analytic_solution} for the derivations).

As mentioned earlier, we divide each l.o.s into small 1D cells and assume a uniform dust density within each cell. The cells are set to have the same size as the typical separation between input stars. The attenuation towards each of these stars is the integral of these 1D dust cells which is then approximated as the sum. Then they are connected in 3D space using the Gaussian process prior as explained above. As discussed in \cite{Rezaei_Kh_17}, the Gaussian prior contains some ``hyperparameters'': One is the correlation length, $\lambda$, which needs to be a few times the cell sizes in order to connect them in 3D space. Another hyperparameter, ${\theta}$, (units ${pc}^{-2}$) is calculated based on the variance in the input density distribution \citep[see equation 16 in][]{Rezaei_Kh_17}. The third hyperparameter is the mean of the prior, $\meanrho$ in equation \ref{eqn:gpJn}. In this work we assume a common (${\rho}_{\mu}$) mean for the entire data set which represents the average dust density in the sample. As our dust density is just attenuation per unit distance, we calculate ${\rho}_{\mu}$ for each set of input stars by dividing the l.o.s attenuation towards each star by its distance to get an average density per star, then set the ${\rho}_{\mu}$ to the mean of these values and $\meanrho$ = ${\rho}_{\mu} \textbf{1}$. In the next section we describe our input data and their corresponding hyperparameters for calculating dust densities.

\section{APOGEE Data}\label{data}
\begin{figure} 
\begin{center}
\includegraphics[width=0.50\textwidth, angle=0]{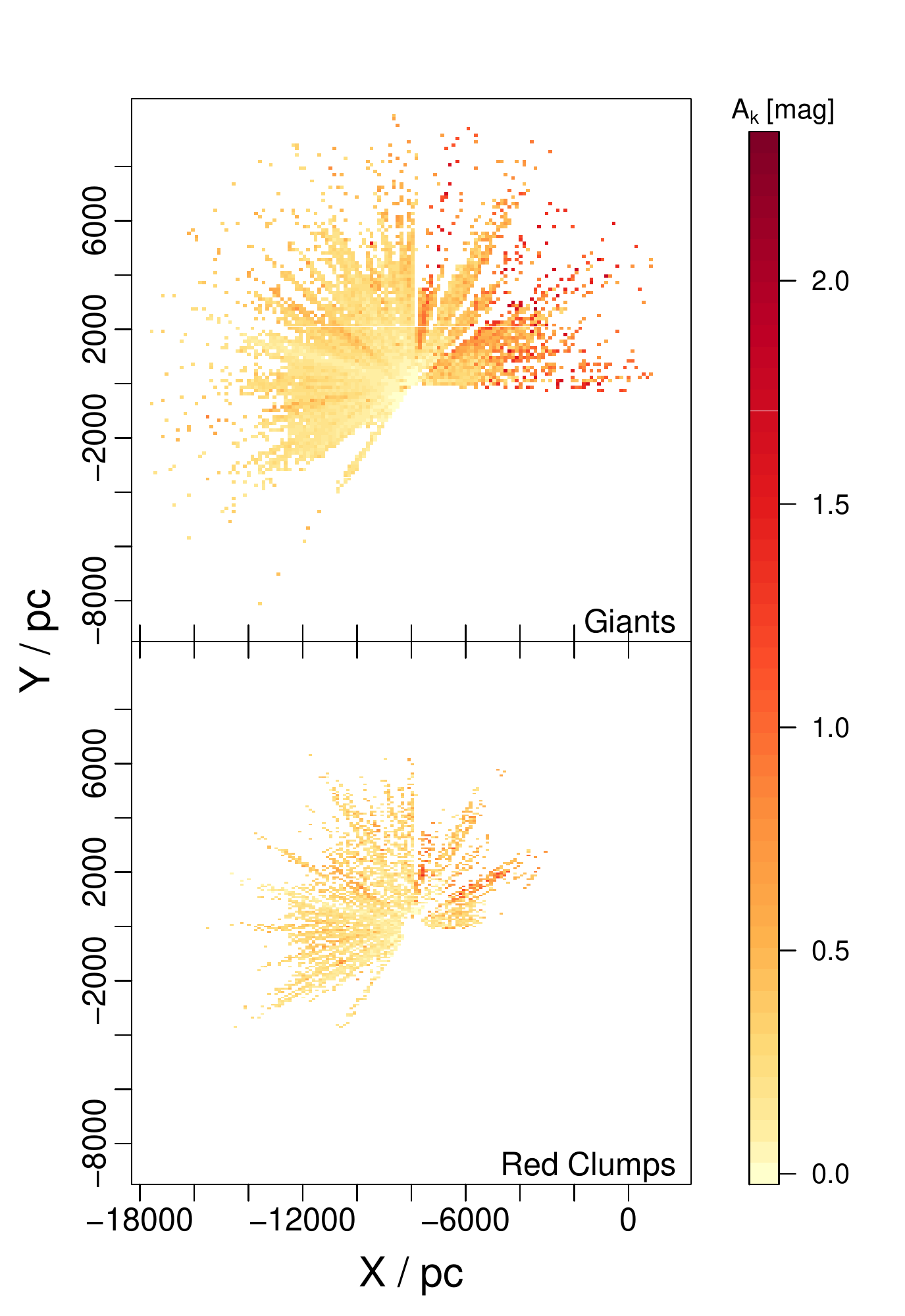}
\caption{Distribution of input stars in the X-Y plane ($\pm$100 pc in Z). The sun is at (-8000, 0) and the Galactic centre is at (0,0). The top panel shows giant stars and the bottom panel shows red clump stars. The colour represents their l.o.s K-band extinction. Most giants can probe out to about 7 kpc, while red clump stars probe to about 5 kpc on average.}
\label{fig:data}
\end{center}
\end{figure}

In this work we use data from APOGEE-2 \citep{Blanton17, Majewski17, Abolfathi17}, a near-infrared high-resolution spectroscopic survey targeting bright stars \citep{APOGEE, Zasowski13}. As the survey operates in the near-infrared, the effects of extinction are about an order of magnitude lower than at optical wavelength, enabling it to observe stars in the highly obscured regions of the Galactic disk and towards the Galactic centre.

We select giants from APOGEE DR14 \citep{Majewski17, Abolfathi17} using their $\rm log\,g$ information (0.5 < $\rm log\,g$ < 3.5). We estimate distances and colour excess ( E(J -- K), as described below) using the stellar parameters ($\rm T_{eff}$, $\rm log\,g$ and [M/H]) together with isochrones from stellar evolution models \footnote{http://stev.oapd.inaf.it/cgi-bin/cmd} \citep[PARSEC,][]{Tang14, Chen15}. The stellar parameters were determined by the APOGEE Stellar Parameters and Chemical Abundances Pipeline \citep[ASPCAP,][]{Garcia16}.

For each star, we look for the closest point in the $\rm T_{eff}$ vs. $\rm log\,g$ plane for the corresponding isochrone at a given metallicity, taking into account the uncertainties in the stellar parameters. We use the calibrated stellar parameters (PARAM) which were calibrated using a sample of well-studied field and cluster stars as well as stars with asteroseismic stellar parameters \citep{Holtzman15}.
Each star in the APOGEE sample has 2MASS J, H, and $K_{s}$ magnitudes. In the isochrone grid we find the absolute magnitudes $M_{J}$ , $M_{H}$, and $M_{K_{s}}$ corresponding to the stellar parameters. The colour excess E(J -- $K_{s}$) is calculated as E(J -- $K_{s}$) = J -- $K_{s}$ -- ($M_{J}$ -- $M_{K_{s}}$).  To convert E(J -- K) to $A_{K_{s}}$ we use the extinction law of \cite{Nishiyama09} with $\rm A_{K_{s}} = 0.528 E(J - K_{s})$.

We then calculate the distances as
\begin{alignat}{2}
d = 10^{ 0.2 (K_{s} - M_{K_{s}}) + 5 - A_{K_{s}}}
\label{eqn:dist_M}
\end{alignat}
For a more detailed description see \cite{Schultheis14}. The distance from each star to the nearest point in the isochrone grid in the $\rm T_{eff}$  and $\rm log\,g$ dimension, together with the individual uncertainties $\rm \sigma_{Teff}$ and $\rm \sigma_{log\,g}$ (from the ASPCAP pipeline), gives the
distance uncertainty for each star. As shown in \cite{Schultheis14} the median uncertainty is approximately 30\%.

Apart from giants, APOGEE also targets red clump (RC) stars. \cite{Bovy14} introduced a new method to select these RC stars from APOGEE data based on their position in the colour--metallicity--surface-gravity--effective-temperature space. Because of the narrowness of the RC locus, distances to these stars can be estimated with an accuracy of 5\% -- 10\% \citep{Bovy14}. The extinctions for this sample are calculated by \cite{Bovy14} using Rayleigh Jeans Colour Excess method \citep[RJCE;][]{Majewski11} which relies on the fact that near- to mid- infrared colours are almost constant for all stars. Therefore the change in the colour of a star is due to interstellar extinction. The catalogue of RC stars is available from APOGEE DR14 containing around 30\,000 stars with accurate distance and extinction estimates \citep{Bovy14}, which we use as inputs for our model to map the dust distribution.
\begin{figure*}
\resizebox{\hsize}{!}{\includegraphics[clip=true]{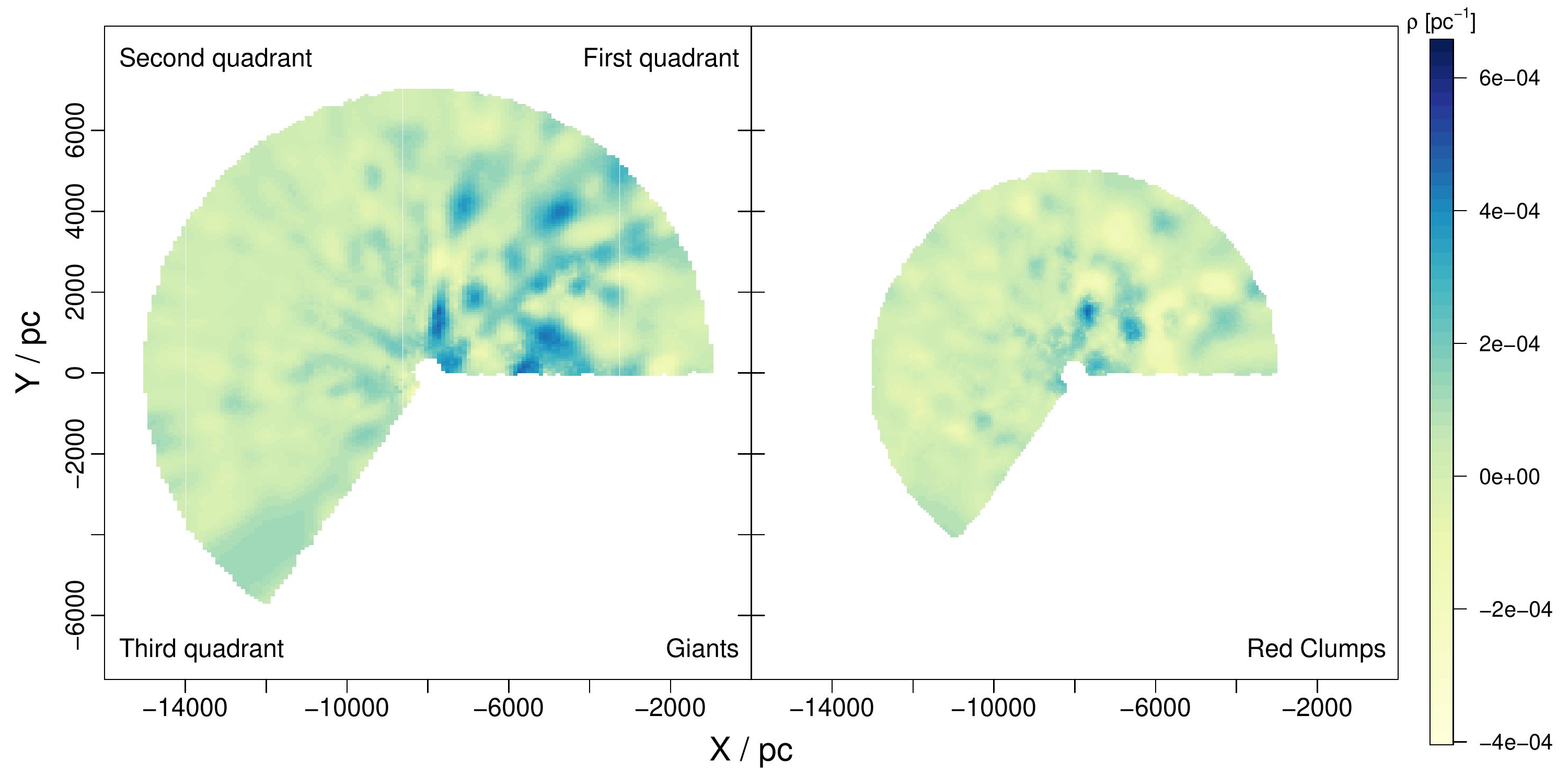}}
\caption{2D image of the 3D map of the dust distribution in the Galactic disk ($\pm$100 pc in the Z direction) using giants (left panel) and RC stars (right panel). The sun is at (-8000, 0) and the Galactic centre at (0,0). The colour shows the mean of the dust density predictions over the column through the disk (${pc}^{-1}$).
}
\label{fig:maps}
\end{figure*}
\begin{figure*}
\resizebox{\hsize}{!}{\includegraphics[clip=true]{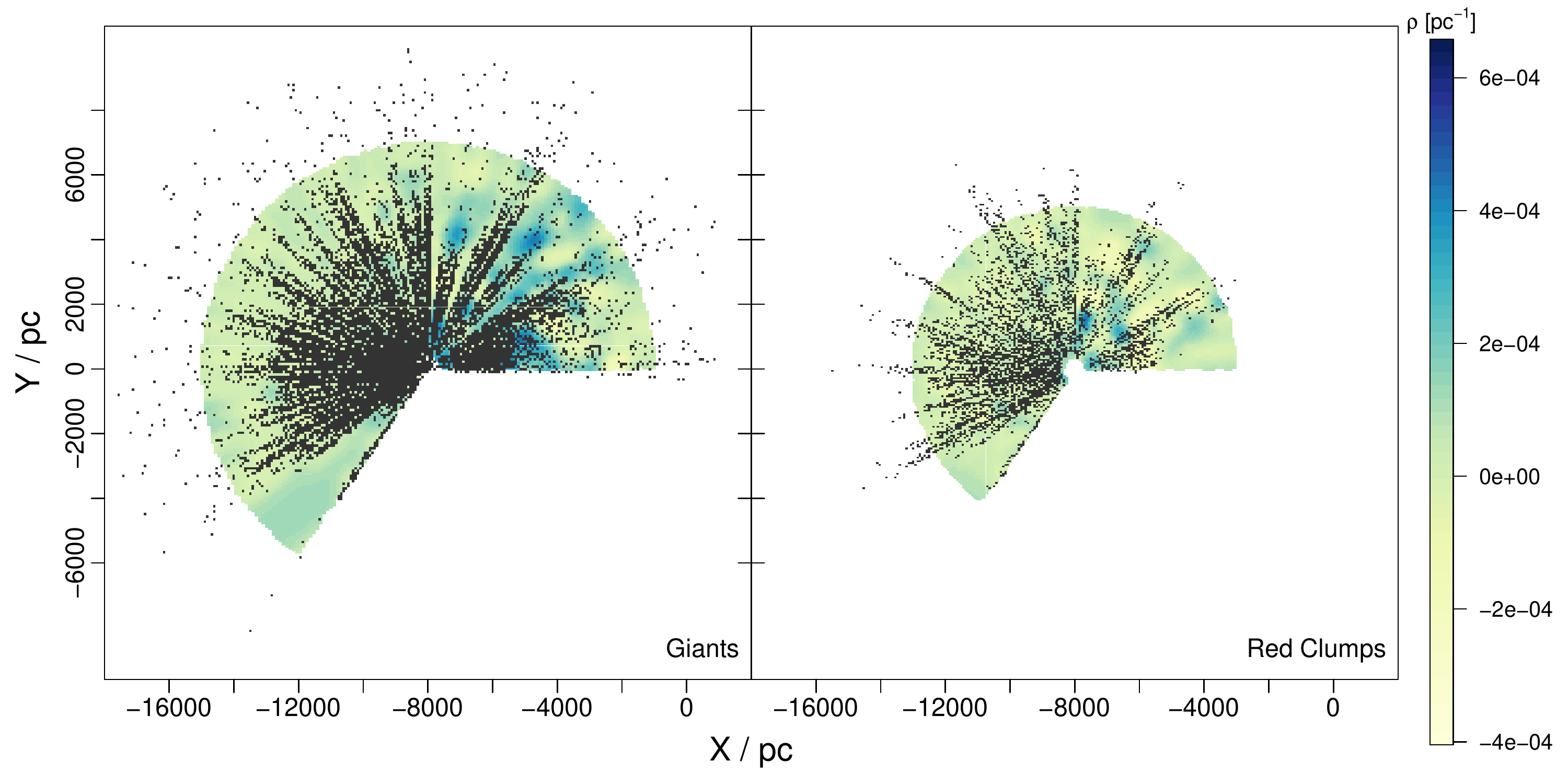}}
\caption{As Fig. \ref{fig:maps} but now overplotting as black points the locations of the stars used to derive the underlying dust map. The giant sample extends to larger distances than the RC stars.
}
\label{fig:map_withinput}
\end{figure*}

From both the giant and RC catalogues we select targets within Z=100 pc above and below the Galactic mid-plane. The RC catalogue provides the Galactic Z values assuming the Sun is 8 kpc from the Galactic centre and 25 pc above the Galactic mid-plane \citep{Bovy14}. We make the same assumption and calculate the Galactic Z for the giants to select stars in the disk of $\pm$100 pc. Also, to be consistent with RC distance precision, from our giant sample we only select those with fractional distance uncertainties less than 0.05. This leaves us with about 5\,000 stars from the RC catalogue and about 16\,000 giants. Figure \ref{fig:data} shows both samples: the giant sample probes larger distances and gives higher extinction measurements, in particular in the first quadrant and towards the Galactic centre.

\begin{figure*}
\resizebox{\hsize}{!}{\includegraphics[clip=true]{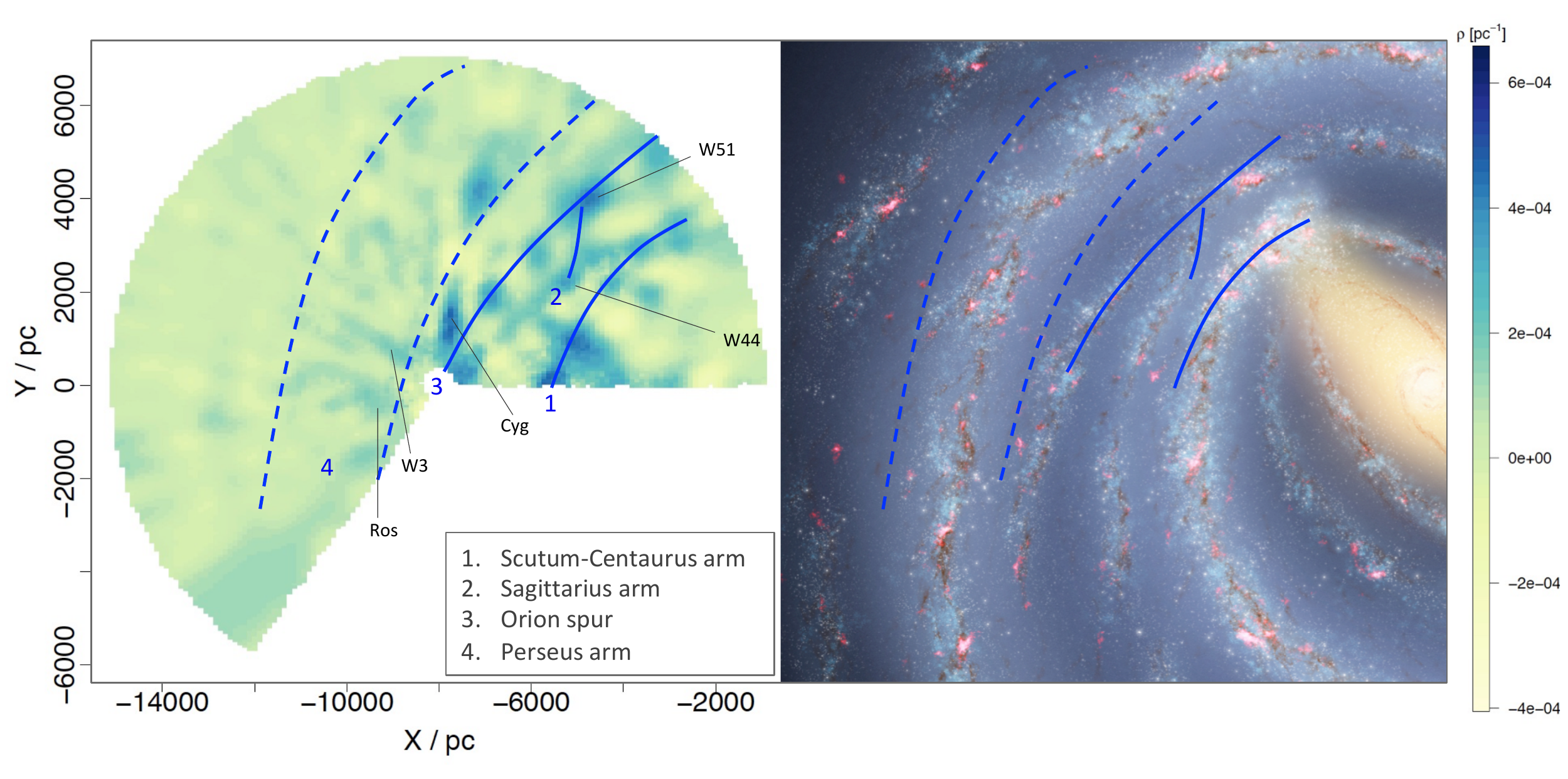}}
\caption{Left panel: our dust density predictions as in Fig. \ref{fig:maps} (left panel), but now over-plotting with blue lines the approximate locations of the arms as we deduce them from this dust map. The dashed lines show an area in which relatively high density dust clouds are seen, but which do not lead to as such a distinct pattern as seen for the other three lines. The known giant molecular clouds detected in the map are also labelled. Right panel: our estimated location of the arms (blue lines) from the left panel plotted on top of the Spitzer sketch of the Galactic arms (by Robert Hurt, courtesy of NASA/JPL-Caltech/ESO).
}
\label{fig:spiral}
\end{figure*}
We use each of these samples separately as the input data in our model to infer the dust distribution in the Galactic disk. The typical separation between stars in both RC and giant samples is of the order of 200 pc. We use this as the cell size in the model, and adopt a correlation length of $\lambda=1000$\, pc. The corresponding $\theta$ for giant and RC samples is $9 {\times} {10}^{-9}$ and $6.5 {\times} {10}^{-9}$ ${pc}^{-2}$ respectively (see section \ref{method}). The common mean dust density of the Gaussian process prior (${\rho}_{\mu}$) is computed to be $1.2 {\times} {10}^{-4}$ ${pc}^{-1}$ for both the giants and the RC samples (see section \ref{method}). It is important to note that the correlation length scale, $lambda$, is \textit{not} the resolution limit of our maps: we can probe much smaller scales depending on the distance between stars \citep{Rezaei_Kh_17,Rezaei_Kh18a}. The correlation length is in fact the upper limit on the distance between two points which can still interact, since we use a covariance function with finite support.

\section{Galactic dust map}\label{map}
Figure \ref{fig:maps} shows our dust maps for both the giants and the RC sample. There are many high dust density clouds and structures in both maps. They trace similar structures in the second and third quadrants out to distances where they overlap: there are many dust clouds with relatively high densities (higher than the average mean density of $1.2 {\times} {10}^{-4}$ ${pc}^{-1}$) spread around the area. However, the maps show dramatic differences in the first quadrant. The reason for this can be seen from Fig. \ref{fig:map_withinput} that shows the input data over-plotted on the predicted dust densities: it is clear that the RC sample poorly covers the first quadrant, especially towards the Galactic centre. In this case the posterior will be dominated by the prior. The giant sample, on the other hand, covers much greater distances in general, in particular they cover most of the area in the first quadrant, resulting in better constrained predictions in the first quadrant compared to that of the RC sample.

Another area of interest in Fig. \ref{fig:map_withinput} is the lower left corner of the maps where there is a gap in the input stars in both the RC and giant samples. For close distances where the stars on both sides of the gap are within the 1000 pc correlation length, predictions are affected by the neighbouring stars. For more distant points, however, the correlation drops to zero and the predictions would be close to the input mean of the Gaussian process prior. The same criterion applies to the outer regions of the first quadrant in the map produced using RC stars.

We now consider only the map based on the giants, since it extends to greater distances. The high density dust clouds in the first quadrant of this map seem to be lined up to form arc-shaped structures along the expected locations of the spiral arms of the Galaxy. To explore this, we connect the lined-up high density clouds to predict the locations of the arms. As seen in Fig. \ref{fig:spiral}, left, in the first quadrant, we connect three sets of lined-up clouds with blue lines; however, in the second quadrant, the high density clouds seem to be spread widely and do not shape clear lines; therefore, we show borders (as dashed lines) of the area where most of high density clouds are located. Afterwards, we over-plot our predictions on the Spitzer sketch of the Galaxy\footnote{https://www.eso.org/public/images/eso1339e/} (Fig. \ref{fig:spiral} right). It is worthwhile to mention that this artist's impression of the Milky Way is based on the Spitzer observations; the spiral structure in the impression is more qualitatively related to observational data. This is the the reason we choose to compare our spiral arm predictions with this particular map. Other present Galactic spiral arm models are heavily model-dependant, consequently quite different from one another, while the Spitzer map is mainly based on the stellar observations and does not have strong Galactic model assumptions, which makes it similar to our approach.

The locations of our predicted arms in the first quadrant are in relatively good agreement with the Spitzer arm structures, especially at the position of the Orion spur (line 3 in Fig. \ref{fig:spiral}). Parts of the Scutum-Centaurus and Sagittarius arms (arms 1 and 2 in Fig. \ref{fig:spiral}) also match nicely with our inferred dust clouds. The location of the Perseus arm coincides with high density clouds in between our predicted two dashed lines. This is in agreement with the recent finding of \cite{Baba18} which uses data from Gaia DR1 concluding that the Perseus arm is being disrupted. 

In addition to spiral arm structures, our map detects some of the known giant molecular clouds (GMCs) that are labelled in Fig. \ref{fig:spiral}. Literature distances to the W51 GMC ranges from 5 to 8 kpc depending on the method used \citep{Parsons12}. From our map, the distance to W51 GMC is about 5.5 kpc, which is in agreement with the distances obtained by \cite{Sato10} and \cite{Russeil03}. We estimate the distance to the W44 GMC, the dust cloud that seems to be associated with the Sagittarius arm, to be about 3 kpc. There are two nearby clumps of high-density dust at the location of the Cygnus GMC at distances of about 1.5 and 2 kpc. The distance to the main OB associations in Cygnus X has been reported to be about 1.7 kpc \citep{Schneider06}. We also see some moderate density of dust at the expected locations of the W3 and Rosette GMCs at about 2 and 1.5 kpc respectively.

\begin{figure} 
\begin{center}
\includegraphics[width=0.50\textwidth, angle=0]{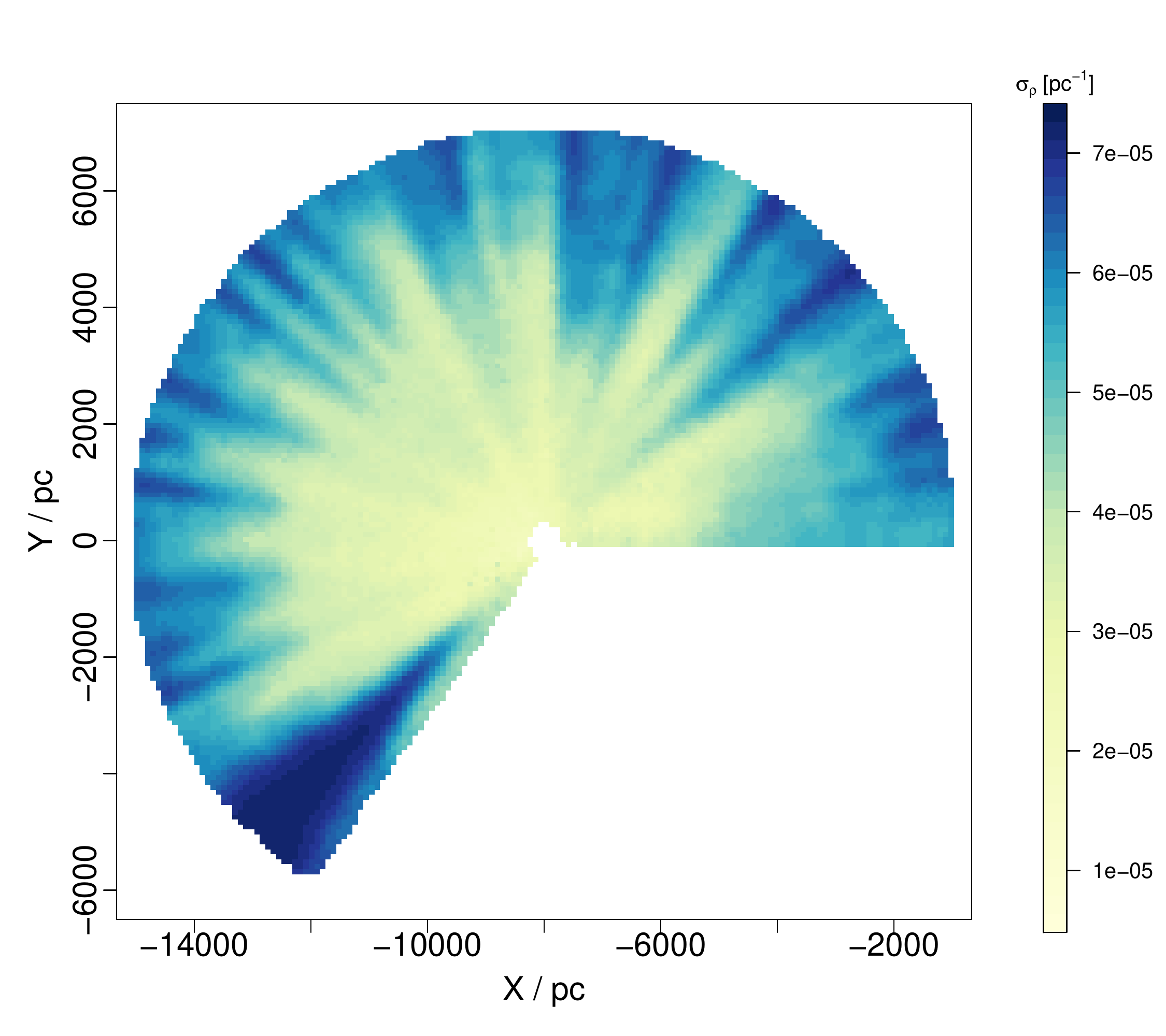}
\caption{Standard deviation of the dust density predictions. The larger uncertainties (dark blue colour) appear at the places not well-populated by the stars.}
\label{fig:sd}
\end{center}
\end{figure}
Figure \ref{fig:sd} shows the model predicted uncertainties. Generally the uncertainties increase as going to larger distances due to the drop in the density of stars in the sample. Comparing this to the input data (Fig. \ref{fig:map_withinput}, left) reveals that regions of larger uncertainty occur where the gaps between the input stars are larger, resulting in the dust density being less well constrained. As can be seen from Fig. \ref{fig:spiral}, we are not able to constrain the location of the Sagittarius arm where it comes close to the sun. This is due to the lack of precise measurements in the direction towards the Galactic centre and lacking data in the southern hemisphere. There is a relatively high density line-shaped area parallel to the Orion spur that seems to be part of the Sagittarius arm. But Fig. \ref{fig:sd} suggests the predictions in this area have relatively high uncertainties and are inferred primarily from their surrounding points. We therefore decided to not include it as part of the arms.

\section{Discussion}\label{discussion}
\begin{figure*}
\resizebox{\hsize}{!}{\includegraphics[clip=true]{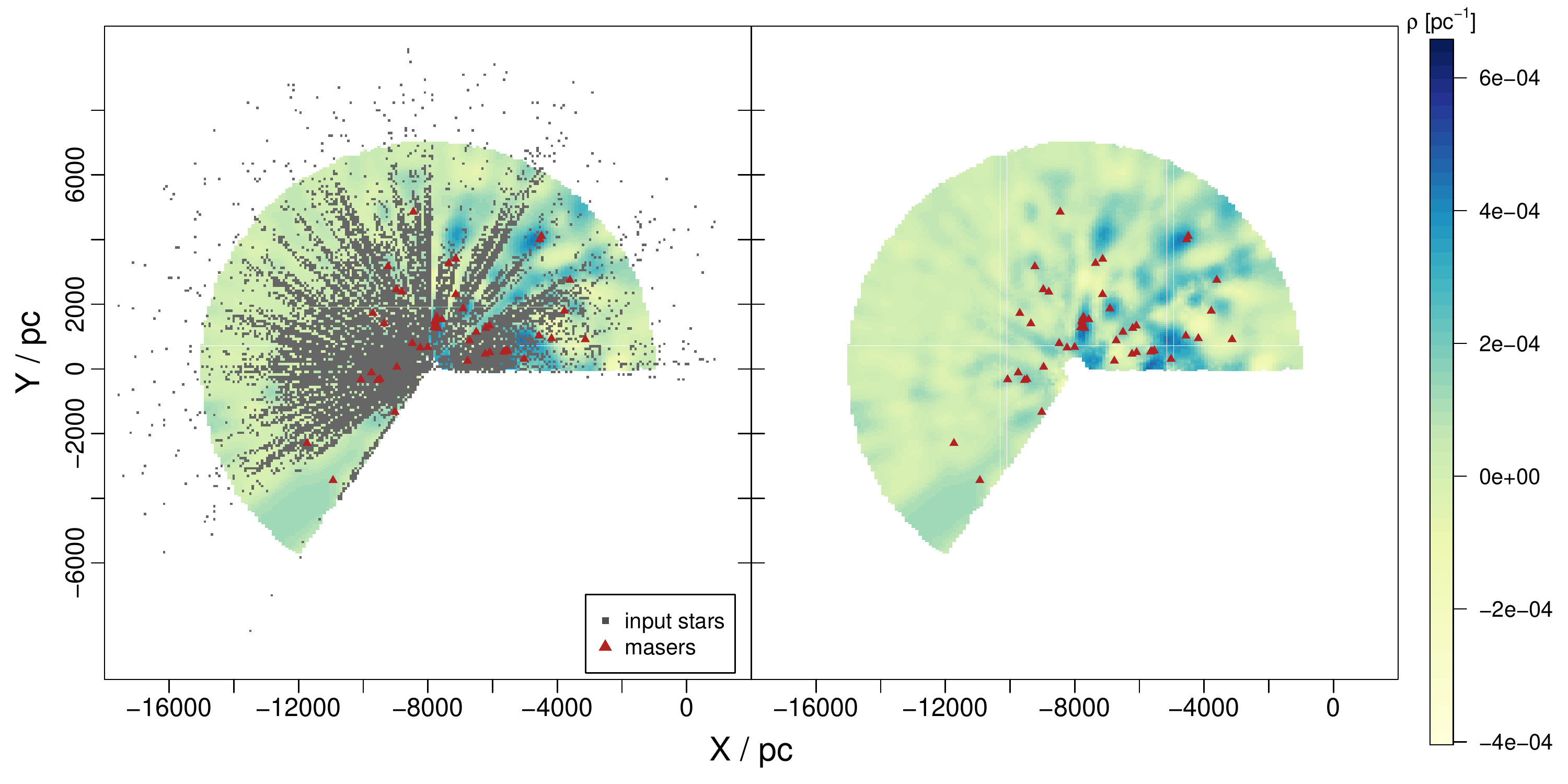}}
\caption{The left panel is the same as the left panel of Fig. \ref{fig:map_withinput}, but now overplotting as triangles the positions of masers from \cite{Reid14} which are within $\pm$100 pc of the disk midplane. Some of the masers show correlation with high density predictions. The right panel is the same, but with the stars removed.
}
\label{fig:map_masers}
\end{figure*}

Our model does not assume any functional form or prior assumption in favour of the spiral arm or Galactic disk structure. Any feature in our map is the outcome of the input data used for inferring the underlying dust densities coupled with the smoothness assumption from the Gaussian process. In order to draw a stronger conclusion for the arm structure of our Galaxy, velocity information from stars could be of great help. Future data from Gaia, 4MOST and WEAVE will provide precise distances, radial velocities and proper motions for millions of stars in the Milky Way which allow us to trace the motion of stars in the Galaxy in order to reveal better the position of the spiral arms. It will be possible then to see whether the arm predictions by the stellar velocities and over-densities coincide with what the dust density probes.

Beyond stellar kinematics, HI and CO observations of gas provide longitude-velocity information which can trace the spiral arms of the Galaxy \citep*[e.g.][]{Dame01}. Since the molecular gas and dust are mixed in the ISM \citep[e.g.][]{Tielens05, Corbelli12}, they are expected to trace the same arm structure. However, the main challenge of using gas velocity information is that their distance estimates (kinematic distances) have large uncertainties , so the sources cannot be precisely located. There have been various studies trying to overcome this issue and determine distances to the star forming regions \citep{Wienen15, Whitaker17} thought to be associated with the spiral arms. One of the major works in this aspect is that of \cite{Reid14} who use trigonometric parallaxes and proper motion of masers associated with high-mass star forming regions that trace the spiral arms in the Milky Way \citep{Reid14}. From their sample we select those masers that are within our probed volume (out to 7 kpc in distance and $\pm$100 pc in the Z direction) and over-plot them on our dust map (Fig. \ref{fig:map_masers}). Some of the maser locations match inferred dust density clouds we find, especially those located in regions well-populated by the input stars. Some others, on the other hand, appear where little dust is inferred. This may be due to the fact that some regions are poorly covered as a result of APOGEE's target selection \citep{Zasowski13}. A more likely reason is the nature of the masers: they cover much smaller physical scales (less than 1000 AU) than the resolution of our map (200 pc scale here, which is related to the typical separation of stars in the sample: see section\ref{data}). The supposed high dust density they trace only extends over a small volume, so makes only a small contribution to the average over a larger volume.
\begin{figure} 
\begin{center}
\includegraphics[width=0.50\textwidth, angle=0]{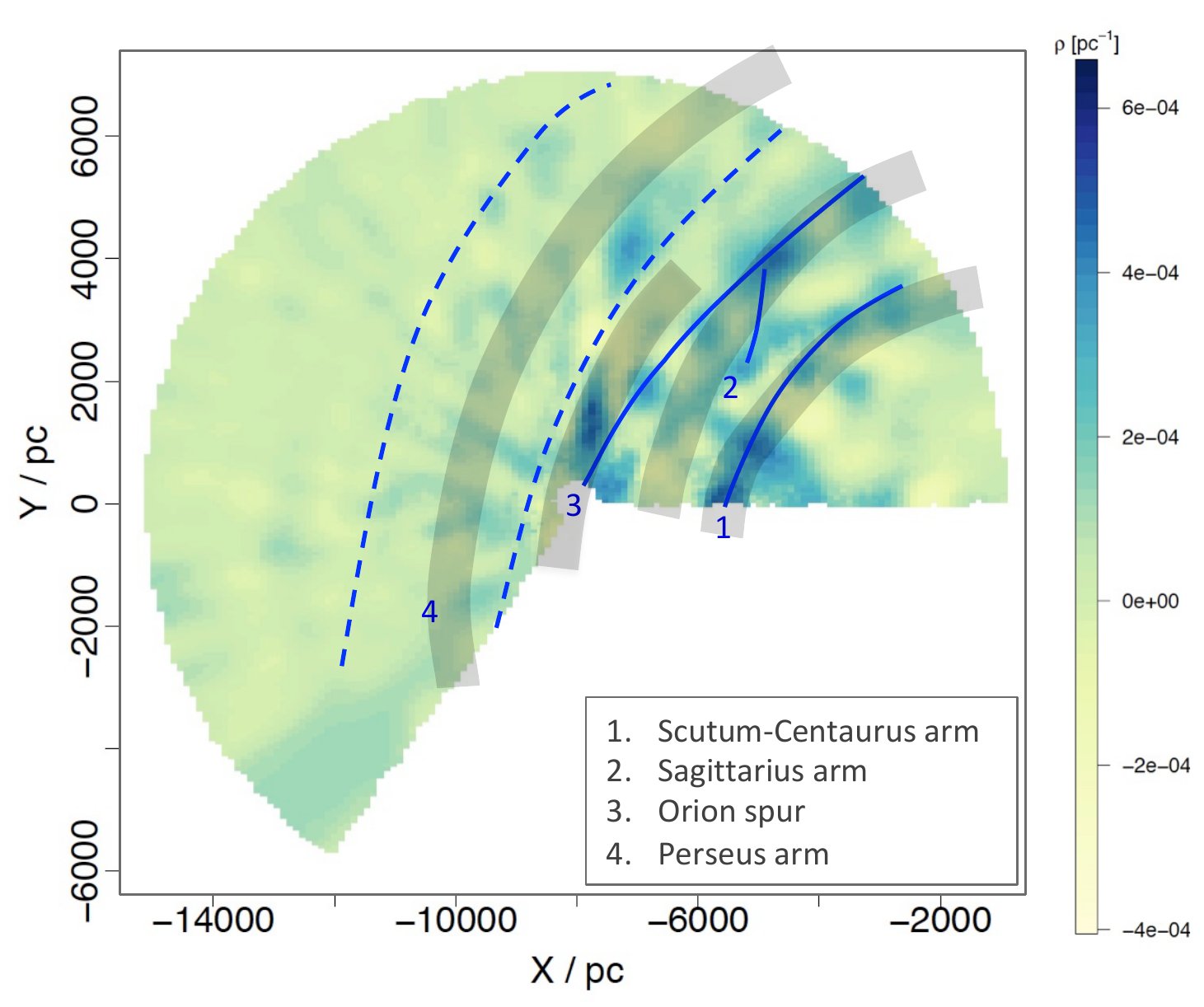}
\caption{As Fig. \ref{fig:spiral} left, but overplotting spiral arms model from \cite{Reid14} as grey shaded curves. The width of the shaded curves is equal to 2$\sigma$ uncertainty in the arm fitting \citep[1$\sigma$ on each side of the central curve, ][]{Reid14}.}
\label{fig:twoarms}
\end{center}
\end{figure}

Figure \ref{fig:twoarms} is the same as the left panel of Fig. \ref{fig:spiral} but additionally shows the spiral arms from the \cite{Reid14} model. Our predicted locations for the Scutum-Centaurus arm and the Perseus arm match that of \cite{Reid14} quite well. In contrast, our predictions for the Sagittarius arm and the Orion spur (local arm) do not match, except for the part of the Sagittarius arm that merges with the Orion spur. These differences could be due to the observational limitations in both cases, i.e. a limited number of lines-of-sights/regions have been observed. It is also important to note that our results are determined without recourse to any prior information on the nature or even the existence of spiral arms or any other Galactic structures. The spiral arm model of \cite{Reid14}, in contrast, contains strong assumptions for the Galactic and arm rotations and structures. Moreover, the assignment of the masers to the spiral arms based on their longitude-velocity measurements is relatively undetermined, especially for the inner arms. Only based on the masers locations in \cite{Reid14}, which represent the locations of the high-density clouds as in our map, it is extremely hard to reach a conclusion about the spiral arms.

Apart from limitations in the precision of the data, there are limitations in our results due to the assumptions of the model. We assume a constant dust density within each cell to relate the measured extinction to the model dust density along observed lines of sight (section \ref{method}). This, together with the separation between stars, limits the resolution of the map and can cause the map to miss smaller structures in the ISM, as discussed above with the masers. We also assume a constant for the mean of the dust in the Gaussian process prior everywhere in our map. This has the advantage of not imposing any spatial dust density prior to the data, therefore the features appear in the map are mainly derived by the input data (i.e. there is no spatial preference for appearance of the high density clouds in the prior), but has the disadvantage of not being very informative. Using a more informative prior, like the mean density derived from an actual map, will better constrain the predictions in the area not well-populated by the data.

As mentioned in \citet{Rezaei_Kh18a}, we do not constrain our dust density predictions or the input extinctions to be positive; as a result, our dust density predictions contain negative values. This is indeed non-physical but is due to the fact that the data is by nature noisy. Both extinctions and distances are measurements which contain uncertainties. Having inconsistent distance-extinction estimates and underestimated uncertainties would cause more/larger negative predictions. In fact, in the presence of the ``good'' data, the predictions are positive \citep[as seen in the results of the simulated data in][]{Rezaei_Kh_17}.

\section{Conclusion}\label{conclusion}
We have presented a map of the Galactic disk using RC stars and giants from APOGEE DR14. This is the first time that such a continuous map of the dust in the disk is presented out to 7 kpc from the Sun. We showed that some of the dust features in our map are possibly associated with spiral arms in our Galaxy. However, our result is limited by the spatial coverage of the input data and observational artefacts due to the APOGEE target selection, plus a limited distance precision of 5\%. Future data from APOGEE south and SDSS-V will be great compliments to the current data by covering the southern hemisphere and providing continuous observation. In addition, the upcoming Gaia DR2 data will be able to cover some of the missing l.o.s in our current APOGEE sample, although the fact that it is an optical survey limits the depth it can probe in dusty regions.

\section*{Acknowledgments}
SRKH would like to thank Henrik Beuther for his constructive discussion and comments. This project is partially funded by the Sonderforschungsbereich SFB\,881 ``The Milky Way System'' of the German Research Foundation (DFG).\\
SDSS-IV is managed by the Astrophysical Research Consortium for the
Participating Institutions of the SDSS Collaboration including the
Brazilian Participation Group, the Carnegie Institution for Science,
Carnegie Mellon University, the Chilean Participation Group, the French Participation Group, Harvard-Smithsonian Center for Astrophysics,
Instituto de Astrof\'isica de Canarias, The Johns Hopkins University,
Kavli Institute for the Physics and Mathematics of the Universe (IPMU) /
University of Tokyo, Lawrence Berkeley National Laboratory,
Leibniz Institut f\"ur Astrophysik Potsdam (AIP),
Max-Planck-Institut f\"ur Astronomie (MPIA Heidelberg),
Max-Planck-Institut f\"ur Astrophysik (MPA Garching),
Max-Planck-Institut f\"ur Extraterrestrische Physik (MPE),
National Astronomical Observatories of China, New Mexico State University,
New York University, University of Notre Dame,
Observat\'ario Nacional / MCTI, The Ohio State University,
Pennsylvania State University, Shanghai Astronomical Observatory,
United Kingdom Participation Group,
Universidad Nacional Aut\'onoma de M\'exico, University of Arizona,
University of Colorado Boulder, University of Oxford, University of Portsmouth,
University of Utah, University of Virginia, University of Washington, University of Wisconsin,
Vanderbilt University, and Yale University.

\bibliographystyle{aa}
\bibliography{Rezaei_Kh._2018_1}

\begin{appendix}
\section{Analytic solution of the integral}\label{sec:analytic_solution}
As explained before, our likelihood function is:
\begin{alignat}{2}
P(\extvecN | \geomat\rhovecJ) \,&=\, \frac{1}{(2\pi)^{N/2}|\likecovN|^{1/2}}  \nonumber\\
   \,&\,\,\quad \mby \,\,\exp\left[ -\frac{1}{2} (\extvecN - \geomat\rhovecJ)\trans \likecovN\inv (\extvecN - \geomat\rhovecJ) \right]  \ ,
\label{eqn:likeN}
\end{alignat}
where $\extvecN$ is the vector of attenuation measurements with covariance $\likecovN$. Using our new Gaussian process prior, the law of marginalization and then applying Bayes theorem, we can write the posterior as
\begin{alignat}{2}
P(\rhoJp | \extvecN) \,&=\, \int_{\geomat\rhovecJ} P(\rhoJp, \geomat\rhovecJ | \extvecN) \, d(\geomat\rhovecJ) \nonumber\\
   \,&=\, \int_{\geomat\rhovecJ} \frac{P(\rhoJp, \geomat\rhovecJ) P(\extvecN | \rhoJp, \geomat\rhovecJ)}{P(\extvecN)} \, d(\geomat\rhovecJ) \nonumber\\
   \,&=\, \frac{1}{P(\extvecN)} \int_{\geomat\rhovecJ} P(\rhoJp, \geomat\rhovecJ) P(\extvecN| \geomat\rhovecJ) \, d(\geomat\rhovecJ) \ .
\label{eqn:rhointA}
\end{alignat}
Both the first (the new Gaussian process prior) and the second (the likelihood) terms are linear functions of $\rhoJp$ which suggests an analytic solution for the integral. For brevity we write equation \ref{eqn:rhointA} as
\begin{equation}
P(\rhoJp | \extvecN) \,=\, \frac{1}{Z} \int_{\geomat\rhovecJ} e^{-\psi/2} \, d(\geomat\rhovecJ)
\label{eqn:rhointB}
\end{equation}
where $Z$ is a normalisation constant, and
\begin{alignat}{2}
\psi \,&=\, \rhovecJp\trans \gpcovJp\inv \rhovecJp +  (\extvecN - \geomat\rhovecJ)\trans \likecovN\inv (\extvecN - \geomat\rhovecJ) \nonumber\\
\,&=\, \rhovecJp\trans \gpcovJp\inv \rhovecJp  +  \extvecN\trans\likecovN\inv\extvecN -  \rhovecJ\trans\geomat\trans\likecovN\inv\extvecN + \rhovecJ\trans\geomat\trans\likecovN\inv\geomat\rhovecJ \nonumber\\ 
&\,\,\quad - \extvecN\trans\likecovN\inv\geomat\rhovecJ
\label{eqn:psiA}
\end{alignat}
where the third and fifth terms are identical as each term is a scalar and
\begin{alignat}{2}
\rhovecJp \,&=\, \left[ \begin{array}{c} \geomat\rhovecJ - \geomat\meanrho \\ \rhoJp - {\rho}_{\mu} \end{array} \right] \\
\gpcovJp\inv \,&=\, \begin{bmatrix} \mmatJ & \mvecJ \\ \mvecJ\trans & q \end{bmatrix}
\label{eqn:partgpcovinv}
\end{alignat}
where $\mmatJ$ ($N\mby N$ matrix), $\mvecJ$ ($N\mby 1$ vector), and $q$ (scalar) are components of the inverted matrix, and ${\rho}_{\mu}$ is the mean of the dust densities for the Gaussian process prior and is calculated from the input attenuation (see section \ref{method}). We can then write
\begin{alignat}{2}
\,&\rhovecJp\trans \gpcovJp\inv \rhovecJp \nonumber\\ 
&\,=\, \left[ \begin{array}{cc} (\geomat\rhovecJ - \geomat\meanrho)\trans & (\rhoJp - {\rho}_{\mu}) \end{array} \right] 
\begin{bmatrix} \mmatJ & \mvecJ \\ \mvecJ\trans & q \end{bmatrix} 
\left[ \begin{array}{c} \geomat\rhovecJ - \geomat\meanrho \\ (\rhoJp - {\rho}_{\mu}) \end{array} \right] \nonumber\\
 &\,=\, (\geomat\rhovecJ - \geomat\meanrho)\trans\mmatJ(\geomat\rhovecJ - \geomat\meanrho) +  2(\rhoJp - {\rho}_{\mu})\mvecJ\trans(\geomat\rhovecJ - \geomat\meanrho) \nonumber\\
 &\,\,\quad + q(\rhoJp - {\rho}_{\mu})^2 \ .
\end{alignat}
Substituting this into equation \ref{eqn:psiA} and gathering together terms gives
\begin{alignat}{2}
\psi \,&=\, (\geomat\rhovecJ)\trans(\mmatJ + \likecovN\inv)\geomat\rhovecJ + 2((\rhoJp - {\rho}_{\mu})\mvecJ\trans - \extvecN\trans\likecovN\inv \nonumber\\
\,&\,\qquad - (\geomat\meanrho)\trans\mmatJ)\geomat\rhovecJ + ((\geomat\meanrho)\trans\mmatJ(\geomat\meanrho)\nonumber\\
  &\,\,\qquad - 2(\rhoJp - {\rho}_{\mu})\mvecJ\trans(\geomat\meanrho) + \extvecN\trans\likecovN\inv\extvecN + q(\rhoJp - {\rho}_{\mu})^2)
\label{eqn:psiB}
\end{alignat}
This is a quadratic expression in $\geomat\rhovecJ$. The last term is independent of $\rhovecJ$ so can be taken out of the integral, allowing us to write equation~\ref{eqn:rhointB} as
\begin{equation}
\begin{split}
P(\rhoJp | \extvecN) \,=\, \frac{1}{Z} \exp\left[-\frac{1}{2}(\geomat\meanrho)\trans\mmatJ(\geomat\meanrho) -\frac{1}{2}\extvecN\trans\likecovN\inv\extvecN \right] 
\\
\mby \,\, \exp\left[-\frac{1}{2}q(\rhoJp - {\rho}_{\mu})^2 + \mvecJ\trans(\geomat\meanrho)(\rhoJp - {\rho}_{\mu}) \right]
\\
\mby \,\, \int_{\geomat\rhovecJ} \exp\left[-\frac{1}{2}(\geomat\rhovecJ)\trans\rmatJ \geomat\rhovecJ + \bvecJ\trans\geomat\rhovecJ \right] d\geomat\rhovecJ
\label{eqn:rhointC}
\end{split}
\end{equation}
where
\begin{alignat}{2}
\rmatJ \,&=\, \mmatJ + \likecovN\inv  \label{eqn:rmatJdef} \\ 
\bvecJ \,&=\, \mmatJ\trans(\geomat\meanrho) + \likecovN\inv\extvecN - (\rhoJp - {\rho}_{\mu})\mvecJ
\end{alignat}
The integral is a standard one allowing us to write equation \ref{eqn:rhointC} as
\begin{alignat}{2}
P(\rhoJp | \extvecN) \,&=\, \frac{1}{Z} e^{-\phi/2} \hspace*{1em} {\rm where} \nonumber\\
\phi \,&=\, q(\rhoJp - {\rho}_{\mu})^2 - 2 \mvecJ\trans(\geomat\meanrho)(\rhoJp - {\rho}_{\mu}) - \bvecJ\trans\rmatJ\inv\bvecJ
\label{eqn:rhointD}
\end{alignat}
and we have absorbed all factors which do not depend on $\rhoJp$ into the normalization constant.
Substituting for $\bvecJ$ this becomes
\begin{alignat}{2}
\phi \,&=\, {\rm(terms\, free\, of}\, \rhoJp) + (q - \mvecJ\trans\rmatJ\inv\mvecJ)(\rhoJp - {\rho}_{\mu})^2 \,+ 
 \nonumber\\
  &\,\,\qquad 2(\extvecN\trans\likecovN\inv\rmatJ\inv\mvecJ + \mvecJ\trans\rmatJ\inv\mmatJ\trans\geomat\meanrho - \mvecJ\trans\geomat\meanrho)(\rhoJp - {\rho}_{\mu}) \ .
\end{alignat}
The first parenthesis contains terms which do not depend on $\rhoJp$ so can be absorbed into the normalization constant. Putting this into equation \ref{eqn:rhointD} gives us
\begin{equation}
P(\rhoJp | \extvecN)  \,=\, \frac{1}{Z} \exp \left[ -\frac{1}{2}\alpha (\rhoJp - {\rho}_{\mu})^2 - \beta (\rhoJp - {\rho}_{\mu}) \right]
\label{eqn:rhointE}
\end{equation}
where
\begin{alignat}{2}
\alpha \,&=\, q - \mvecJ\trans\rmatJ\inv\mvecJ \nonumber \\
\beta \,&=\, \extvecN\trans\likecovN\inv\rmatJ\inv\mvecJ + \mvecJ\trans\rmatJ\inv\mmatJ\trans\geomat\meanrho - \mvecJ\trans\geomat\meanrho \ .
\label{eqn:auxtermsA}
\end{alignat}
By completing the square in the exponent we see that
\begin{equation}
P(\rhoJp | \extvecN) \,=\, \sqrt{\frac{\alpha}{2\pi}} \exp \left[   -\frac{\alpha}{2}\left(\rhoJp + \frac{\beta}{\alpha} - {\rho}_{\mu}\right )^2 \right] \ .
\label{eqn:rhointF}
\end{equation}
which is a Gaussian with mean $-\beta/\alpha + {\rho}_{\mu}$ and variance $1/\alpha$.

\end{appendix}

\end{document}